\documentclass{ieeeaccess}
\usepackage{cite}
\usepackage{amsmath,amssymb,amsfonts}

\usepackage[pdftex]{graphicx,graphics}
\usepackage{lscape,tipa}
\usepackage{multirow,array,bm,enumerate}
\usepackage{amssymb,epsfig,amsmath,cite,comment,hyperref,verbatim}
\usepackage{textcomp}
\usepackage{url}
\usepackage{color}
\usepackage{caption}
\usepackage[dvipsnames]{xcolor}

\def\BibTeX{{\rm B\kern-.05em{\sc i\kern-.025em b}\kern-.08em
    T\kern-.1667em\lower.7ex\hbox{E}\kern-.125emX}}
\begin{document}
\history{Date of publication November 8, 2021, date of current version November 16, 2021.} 
\doi{10.1109/ACCESS.2021.3126280}

\title{Formant Tracking using Quasi-Closed Phase Forward-Backward Linear Prediction Analysis and Deep Neural Networks}

\author{\uppercase{Dhananjaya Gowda *$^{+}$}\authorrefmark{1}, \IEEEmembership{Member, IEEE}, \uppercase{Bajibabu Bollepalli*$^{+}$}\authorrefmark{2}, \IEEEmembership{Member, IEEE}, \uppercase{Sudarsana Reddy Kadiri*}\authorrefmark{3}, \IEEEmembership{Member, IEEE},
 \uppercase{Paavo Alku\authorrefmark{3}},
 \IEEEmembership{Fellow, IEEE}}
 \address[1]{Samsung Research South Korea, E-mail: njaygowda@gmail.com.}
 \address[2]{Amazon UK, E-mail: bajibabu7@gmail.com.}
 \address[3]{Dept. of Signal Processing and Acoustics, Aalto University, FI-00076 Espoo, Finland (\{sudarsana.kadiri;paavo.alku\}@aalto.fi)}

\tfootnote{$*$ Equal contribution.$+$The authors were with Aalto University when this work was carried out. This study was partly funded by the Academy of Finland (Project No. 313390).}

\corresp{Corresponding author: Sudarsana Reddy Kadiri (e-mail: sudarsana.kadiri@aalto.fi).}

\begin{abstract}
Formant tracking is investigated in this study by using trackers based on dynamic programming (DP) and deep neural nets (DNNs). Using the DP approach, six formant estimation methods were first compared. The six methods include linear prediction (LP) algorithms, weighted LP algorithms and the recently developed quasi-closed phase forward-backward (QCP-FB) method. QCP-FB gave the best performance in the comparison. Therefore, a novel formant tracking approach, which combines benefits of deep learning and signal processing based on QCP-FB, was proposed. In this approach, the formants predicted by a DNN-based tracker from a speech frame are refined using the peaks of the all-pole spectrum computed by QCP-FB from the same frame. Results show that the proposed DNN-based tracker performed better both in detection rate and estimation error for the lowest three formants compared to reference formant trackers. 
Compared to the popular Wavesurfer, for example, the proposed tracker gave a reduction of 29\%, 48\% and 35\% in the estimation error for the lowest three formants, respectively.
\end{abstract}



\maketitle

\section{Introduction}
\label{sec:intro}
Estimation and tracking of formant frequencies is an important research topic in several areas of speech science and technology \cite{Assmann1995, Welling1998, Smit2012, Yoo2015, Rita2016,kathania2020study}. 
During the past few decades, many techniques have been proposed for formant tracking \cite{praat2001,wavesurfer2000,lideng2007,Mehta2012}.
These algorithms typically consist of two parts, the estimation stage and the tracking stage. In the former, initial estimates of the vocal tract resonances (VTRs) are computed in short frames (e.g., 25 ms) using spectral estimation methods such as linear prediction (LP). In the latter, the formants estimated from individual frames are expressed using contours which cover a longer unit (e.g., word or sentence) \cite{praat2001,wavesurfer2000}.
In addition, estimation and tracking can be done simultaneously using an initial representation of the vocal tract system \cite{lideng2007,Mehta2012}.
In both approaches, accurate estimation of VTRs is an important and necessary computational block.

LP is the most widely used technique to estimate VTRs from speech \cite{makhoul1975} and therefore 
many variants of LP have been proposed (e.g. \cite{GiacobelloASLP2012, magi2009}). 
In formant estimation and tracking, the most popular variants are the autocorrelation and covariance methods \cite{praat2001,wavesurfer2000}. 
The closed phase (CP) analysis is known to improve VTR estimates by avoiding the contribution of the speech samples in the open phase of the glottal cycle thereby decoupling the effect of the trachea more effectively \cite{Yegna1998}. 
CP analysis, however, works better for low-pitched voices which typically have a larger number of samples in the closed phase of the glottal cycle compared to high-pitched voices which might have just a few samples in the closed phase. To reduce problems caused by having a small number of closed phase samples, LP can be computed over multiple neighboring cycles \cite{Yegna1998}.

Weighted linear prediction (WLP) is an all-pole modeling method based on temporally weighting the prediction error  \cite{magi2009,Mizoguchi1982,ChinHui1988,ma1993,PaavoJASA2013,Manu2014,gowda2020time}. 
Temporal weighting of the prediction error has been shown to be beneficial in computing vocal tract models which are robust with respect to noise and the selection of analysis window as well as the biasing effect of high fundamental frequency. 
Formant estimation of high-pitched vowels was studied using WLP in \cite{PaavoJASA2013} by developing a simple weighting function, called the attenuated main excitation (AME) function, to downgrade the strong effect of the glottal source in the computation of the vocal tract model. 
Based on \cite{PaavoJASA2013}, the quasi-closed phase (QCP) method was proposed for glottal inverse filtering (GIF) in  \cite{Manu2014}. In QCP, a more generalized AME-type of weighting function 
is used. 
Recently, a new formant estimation method based on QCP, called quasi-closed phase forward-backward (QCP-FB) LP analysis, was proposed in \cite{QCPFB_JASA}. QCP-FB combines two approaches: (1) QCP analysis in which the residual is temporally weighted, and (2) forward-backward (FB) analysis in which the number of samples is increased in LP by using two prediction directions simultaneously. In addition, WLP methods have been proposed recently based on using stochastic approaches in the computation of the weighting function \cite{Achuth2019}.

In this article, formant tracking is studied by investigating different all-pole modeling methods in formant estimation. The all-pole formant estimation methods are used with two formant tracking approaches, a dynamic programming (DP) -based approach and a deep neural net (DNN) -based approach. As the first part of the study, six different LP-based and WLP-based formant estimation methods are compared in formant tracking using a DP-based tracker. The novelty of this part is in studying how the potential new method, QCP-FB, which was investigated solely in formant $estimation$ in \cite{QCPFB_JASA}, works in formant $tracking$. In the second part of the study, two most potential all-pole modeling methods from the first part are used with a modern DNN-based tracker by proposing a novel formant tracking approach, which combines benefits of the data-driven deep learning approach and benefits of the model-driven all-pole modeling approach. In this novel tracking approach, the formants, which are predicted by the DNN from a given speech frame, are refined using the spectral peaks, which are indicated by the spectrum, which is computed from the same frame with a model-based parametric all-pole spectral estimation method. Altogether five known formant trackers (Wavesurfer ~\cite{wavesurfer2000}, PRAAT \cite{praat2001}, MUST \cite{Mustafa2006}, KARMA \cite{Mehta2012}, and Deep Formants \cite{Dissen2019}) are used as reference methods in this study.

{
The contributions of the study are as follows:
\begin{itemize}
    
    \item The potential new formant estimation method, QCP-FB, is evaluated in formant $tracking$ and its performance is compared with existing LP-based and WLP-based formant estimation methods using a DP-based tracker.
    
    \item  A novel formant tracking technique is proposed by combining the data-driven DNN-based approach and the model-driven all-pole approach. In this technique, the formants predicted from a speech frame by a DNN are refined using the spectral peaks that are extracted from an all-pole model, which is computed from the same frame.
    
    

    \item A systematic investigation is carried out by comparing the novel formant tracking method described above with five reference formant trackers (Wavesurfer ~\cite{wavesurfer2000}, PRAAT \cite{praat2001}, MUST \cite{Mustafa2006}, KARMA \cite{Mehta2012}, and Deep Formants \cite{Dissen2019}).
    
\end{itemize}}

The paper is organised as follows. The QCP-FB method, which was introduced as a new formant estimation method recently in \cite{QCPFB_JASA}, is first described in section II. The other formant estimation methods and the formant trackers used in the study are described in section III. The results of the formant tracking experiments are reported in section IV. Finally, conclusions are drawn in section V.

\section{Quasi-closed phase forward-backward analysis}
\label{sec:fblp}
The traditional formulation of LP is based on forward prediction in which the current speech sample is predicted from the past $p$ samples.
It is, however, also possible to use 
backward prediction in which the current sample is predicted from the future $p$ samples.
The filter coefficients computed using forward and backward predictions are inter-convertible, and therefore they do not carry any additional information when computed separately. However, by simultaneously using both backward and forward prediction, a prediction model different from that of traditional LP is obtained by using forward-backward (FB) analysis, where the current sample is predicted based on past and future samples using a common set of $p$ coefficients.
The combined error to be minimized is given by
\begin{gather}
{\cal E}={\cal E}^f+{\cal E}^b,
\end{gather}
\begin{gather}
\text{where}\quad{\cal E}^f=\sum_{n}{\left(x_n + \sum_{k=1}^{p}{a_k x_{n-k}}\right)^2}
\end{gather}
\begin{gather}
\text{and}\quad{\cal E}^b=\sum_{n}{\left(x_n + \sum_{k=1}^{p}{a_k x_{n+k}}\right)^2}
\end{gather}
denote the forward and backward errors, respectively, $x_n$ denotes the current speech sample, and $a_k$ denotes the prediction coefficients.
The prediction coefficients can be computed by minimizing the combined error ($\partial{\cal E}/\partial{a_i}=0,\enskip 1\le i\le p$) which results in the following normal equations
\begin{gather}
\sum_{k=1}^{p}c_{i,k}a_k=-c_{i,0}, \quad 1\le i\le p
\end{gather}
\begin{gather}
\text{where} \enskip c_{i,k}=\sum_{n}x_{n-i}x_{n-k} + \sum_{n}x_{n+i}x_{n+k}.
\end{gather}

Previous studies have shown that FB analysis reduces the dependency of spectral estimates on the initial sinusoidal phase, 
shifting of frequency estimates due to additive noise and the so called line-splitting problem 
(see \cite{QCPFB_JASA} for a review).
The line-splitting problem refers to obtaining spectral models which show a single sinusoidal component incorrectly as two distinct peaks. By taking advantage of FB analysis, two benefits are achieved: (1) the estimated spectral peak locations are less sensitive to the window position, and (2) the combination of the two prediction directions gives more samples to compute correlations for the given frame. 

Quasi-closed phase forward-backward (QCP-FB) analysis involves the use of FB analysis within the framework of QCP in order to combine the benefits of both techniques. The resulting method imposes the temporal QCP weighting function $w_n$, defined by \cite{Manu2014}, on the forward and backward errors individually. The combined error to be minimized is given by
\begin{gather}
{\cal F}={\cal F}^f+{\cal F}^b,
\label{eqn:cerror}
\end{gather}
\begin{gather}
\text{where}\quad{\cal F}^f=\sum_{n}{w_n\left(x_n + \sum_{k=1}^{p}{a_k x_{n-k}}\right)^2}
\label{eqn:wtfn1}
\end{gather}
\begin{gather}
\text{and}\quad{\cal F}^b=\sum_{n}{w_n\left(x_n + \sum_{k=1}^{p}{a_k x_{n+k}}\right)^2}
\label{eqn:wtfn2}
\end{gather}
are the weighted forward and backward errors, respectively.
The resulting normal equations are given by
\begin{gather}
\sum_{k=1}^{p}d_{i,k}a_k=-d_{i,0}, \quad 1\le i\le p
\label{eqn:norm}
\end{gather}
\begin{gather}
\text{where} \enskip d_{i,k}=\sum_{n}w_nx_{n-i}x_{n-k} + \sum_{n}w_nx_{n+i}x_{n+k}.
\label{eqn:norm1}
\end{gather}
{Appropriate choice of range for the variable $n$ results in the autocorrelation or covariance methods for QCP-FB. 

QCP-FB is used in formant tracking in the current study and it is expected to show improved performance compared to existing formant tracking methods due to the following two main reasons. First, FB analysis helps to improve the formant estimation by providing more samples for prediction, and by reducing the problems of window positioning and line splitting. Second, QCP analysis exploits the WLP framework of sample selective prediction by designing a temporal weighting function that gives more emphasis on closed phase regions and deemphasizes the open phase as well as the region immediately after the main excitation. This results in more accurate closed phase estimates of the vocal tract system with a reduced influence from the glottal source.} 


\section{Formant trackers}
Several formant tracking algorithms have been proposed in the literature \cite{praat2001,wavesurfer2000,lideng2007,Mehta2012,Dissen2019}. It is worth emphasising that a formant tracking algorithm will most likely show varying performance when combined with different formant estimation methods and this makes it difficult to compare different tracking algorithms. In principle, most of the tracking algorithms can be combined with any formant estimation method. Therefore, formant tracking is studied in this paper using trackers which are based on both DP and DNN. 

\subsection{DP-based formant trackers}
Using the DP-based tracking algorithm proposed in \cite{wavesurfer2000}, formant tracking performance was investigated by comparing six different formant estimation methods that all use all-pole modeling. These methods, listed in Table~\ref{tab:ftrack1}, 
are as follows: (1) conventional LP based on the autocorrelation method (LP-ACOR), (2) conventional LP based on the covariance method (LP-COV), (3) LP based on forward-backward prediction and the covariance method (LP-FBCOV), (4) QCP analysis based on the autocorrelation method (QCP-ACOR), (5) QCP analysis based on the covariance method (QCP-COV) and (6) QCP analysis based on forward-backward prediction and the covariance method (QCP-FBCOV). All these methods were computed using a frame length of 25 ms, a frame shift of 10 ms and an all-pole model order $p$=12. Speech signals, sampled using 8 kHz, were pre-emphasised using an FIR filter ($P(z)=1-0.5z^{-1}$). In the autocorrelation methods, the Hamming window was used. In the covariance methods, the rectangular window was used. The peaks in the spectrum were detected by convolving the spectrum using a Gaussian derivative window with a width of 100 Hz and picking the negative zero-crossings. Five most energetic peaks of the spectrum were selected as the formant candidates. A verbatim MATLAB implementation of the tracking algorithm \cite{wavesurfer2000} was used to track the best four contours from the underlying formant candidates estimated by the all-pole methods.

\subsection{DNN-based formant trackers}
In order to study the possible limitations of the DP-based tracker, a deep neural network (DNN) -based formant tracker was developed as an alternative. A simple four-layer feed-forward DNN was used to capture the nonlinear mapping between the spectrum and the formant frequencies.
The DNN had 300 units with tangent-hyperbolic activation in each of the three hidden layers~\cite{Goodfellow2016}.
The input dimension of 143 units corresponded to 13 RASTA-PLP \cite{Hynek1990} cepstral coefficients with an 11-frame neighborhood, and the three linear output units corresponded to the first ($F_1$), second ($F_2$) and third ($F_3$) formant to be predicted.

A common input feature was deliberately used to have a common baseline performance, and to study the incremental improvement provided by different spectrum estimation methods when used for refinement.
300 utterances from the train subset of the VTR-TIMIT database ~\cite{lideng2006} were used to train the models.
Mean square error between the estimated and actual formant values was used as the objective function.
All parameters of the network were initialized randomly.

The stochastic gradient descent algorithm with standard backpropagation of error was used to learn the network parameters.
The dropout regularization method was used to prevent overfitting the network.
Input values were normalized to the range of [0.1, 0.9]  and output values were normalized to have zero mean and unit variance. The DNN -based tracker was used in three modes: (1) by predicting the lowest three formants directly, (2) by refining the formants predicted by the DNN by replacing them with the frequencies of the corresponding nearest peaks in the LP-FBCOV spectrum, (3) by refining the formants predicted by the DNN by replacing them with the frequencies of corresponding nearest peaks in the QCP-FBCOV spectrum. (Note that with the model order $p$=12, the LP-FBCOV spectrum and the QCP-FBCOV spectrum can show maximally six peaks). These three trackers will be referred to as DNN, DNN-LP-FBCOV and DNN-QCP-FBCOV, respectively. It is worth emphasizing that the latter two modes combine a data-driven approach and a model-driven approach in formant tracking in a novel way: formants $F_1$--$F_3$ are first predicted using a data-driven $deep$ $learning$ $approach$ from a given frame with the DNN after which the predicted formants are refined using a model-driven $signal$ $processing$ $approach$ using the all-pole spectrum extracted from the frame.


\begin{table*}[h]
\centering
\caption{\label{tab:ftrack1} Formant tracking performance of the DP-based tracker using six all-pole modeling methods in formant estimation and performance of six reference trackers. The numbers in parentheses denote the potential performance of the underlying formant estimation method if any of the five formant candidates is found within the allowed deviation from the ground truth. The results are reported by averaging over of all the 192 utterances of the VTR test database.}
\resizebox{15cm}{4.4cm}{
\begin{tabular}{|c||c|c|c||c|c|c|}
\hline
 & \multicolumn{3}{c||}{FDR (\%)} & \multicolumn{3}{c|}{FEE (Hz)} \\\cline{2-7}
Method ~&~ $F_1$ ~&~ $F_2$ ~&~ $F_3$ ~&~ $\delta F_1$ ~&~ $\delta F_2$ ~&~ $\delta F_3$ ~\\\hline\hline
 \multicolumn{7}{|c|}{All-pole modeling methods} \\\hline
LP-ACOR & 84.3 (92.2) & 72.3 (90.9) & 69.0 (87.4)  & 92 (66) & 296 (135) & 325 (178)\\\hline
LP-COV & 86.0 (92.3) & 75.4 (91.5) & 71.3 (87.9) & 89 (64) & 292 (131)& 319 (174)\\\hline
LP-FBCOV & 86.0 (92.4) & 75.4 (91.9) & 71.3 (88.2) & 89 (64) & 292 (129) & 319 (172)\\\hline
QCP-ACOR & 86.8 (91.6)& 75.5 (91.4)& 71.6 (88.5)& 87 (69)& 292 (132)& 317 (167)\\\hline
QCP-COV & 89.7 (91.6)& 86.1 (91.6)& 79.6 (89.0)& { 73} (69)& 187 (130)& { 228} (165)\\\hline
QCP-FBCOV & { 90.0} (93.4) & 82.1 (93.9)& 77.0 (92.1)& { 73} (63)& 233 (114)& 258 (130)\\\hline
 \multicolumn{7}{|c|}{Reference trackers} \\\hline
PRAAT & 86.0 & 70.0 & 63.1 & 88 & 268 & 340 \\\hline
MUST & 81.1 & { 86.3} & 76.9 & 91 & { 152} & 230 \\\hline
WSURF-0 & 84.1 & 78.2 & 77.3 & 93 & 239 & 245 \\\hline
WSURF-1 & 86.6 & 82.7 & {80.8} & 87 & 223 & {228} \\\hline
KARMA &91.5 &89.4 &74.7 & {62} & 146 & 250 \\\hline 
Deep Formants & 91.7 & 92.3 & {89.7} & 85 & 120 & {143} \\\hline 
\end{tabular}}
\end{table*}

\subsection{Reference formant trackers}
The DP-based and the DNN-based formant tracking algorithms were compared to known formant trackers. These reference trackers include algorithms used in two popular speech analysis tools (Wavesurfer~\cite{wavesurfer2000} and PRAAT~\cite{praat2001}), the adaptive filter bank (AFB) -based formant tracking algorithm (denoted as MUST)~\cite{Mustafa2006}, KARMA (based on Kalman filtering) \cite{Mehta2012}, and Deep Formants (based on DNNs) \cite{Dissen2016,Dissen2019}. Both Wavesurfer and PRAAT use LP analysis followed by DP-based tracking. Wavesurfer was used in two forms corresponding to autocorrelation LP and stabilized covariance LP which are referred to as WSURF-0 and WSURF-1, respectively~\cite{wavesurfer2000}.
The PRAAT algorithm uses the BURG method in LP analysis ~\cite{praat2001}. 
All the algorithms tracked four formants from the top five formant candidates derived from the underlying spectrum at a frame rate of 100 Hz.

\section{Experiments and results}
\subsection{Database and performance metrics}
The formant tracking performance was evaluated using the VTR database, which is one of the most widely used speech databases in the areas of formant estimation and tracking~\cite{lideng2006}. The test data of the database was used for the evaluation. This data consists of 192 utterances (produced by 8 female and 16 male speakers, each pronouncing 8 utterances). The duration of each utterance varies between 2 and 5 s. The ground truth (i.e., formant frequencies) have been derived using a semi-supervised LP--based method \cite{lideng2004}. The values of $F_1$--$F_3$ have been corrected manually using spectrograms. The ground truth values for formants are provided for every 10 ms interval. 
The formant tracking performance was evaluated using two known metrics: the formant detection rate (FDR) and the formant estimation error (FEE) as defined in \cite{QCPFB_JASA}. 
During the performance evaluation, the reference ground truth for each of the lowest three formants was associated with the nearest formant candidate lying within a specified relative ($\tau_p$) and absolute ($\tau_a$) deviations.

The FDR is computed in terms of the percentage of frames for which a hypothesized formant occurs within a specified deviation from the ground truth formant.
The FDR for the $i^{th}$ formant over $M$ analysis frames is computed as
\begin{align}
D_{i} &=  \frac{1}{M}\sum_{n=1}^{M}{I(\Delta F_{i,n})} ,\\
I(\Delta F_{i,n}) &= \left\{\begin{array}{ll} 1 &\quad\text{if} \left({\Delta F_{i,n}}/{F_{i,n}} < \tau_r \quad \& \quad \Delta F_{i,n} < \tau_a\right) \\ 0 &\quad \text{otherwise, }\end{array}\right. \label{eq:fdr}
\end{align}
where $I(.)$ denotes a binary formant detector function. $\Delta{F_{i,n}}=|F_{i,n}-\hat{F}_{i,n}|$ is the absolute deviation of the hypothesized formant  frequency ($\hat{F}_{i,n}$) from the reference ground truth ($F_{i,n}$) at the $n^{th}$ frame for the $i^{th}$ formant.
The FEE is computed in terms of the average absolute deviation of the hypothesized formant from the ground truth formant. The FEE for the $i^{th}$ formant over $M$ analysis frames is computed as
\begin{equation}
R_i=\frac{1}{M}\sum_{n=1}^{M}{\Delta F_{i,n}}.
\end{equation}
The FEE values in conjunction with FDR values give a better sense of the performance of a formant tracker.

\begin{table*}[h]
\centering
\caption{\label{tab:ftrack3} Performance of the DNN-based formant trackers and performance of three reference trackers. The results are reported by averaging over of all the 192 utterances of the VTR test database.}
\resizebox{12cm}{3.3cm}{
\begin{tabular}{|c||c|c|c||c|c|c|}
\hline
 & \multicolumn{3}{c||}{FDR (\%)} & \multicolumn{3}{c|}{FEE (Hz)} \\\cline{2-7}
Method ~&~ $F_1$ ~&~ $F_2$ ~&~ $F_3$ ~&~ $\delta F_1$ ~&~ $\delta F_2$ ~&~ $\delta F_3$ ~\\\hline\hline
 \multicolumn{7}{|c|}{DNN-based formant trackers} \\\hline
DNN & 90.5 & 91.6 & 82.1 & 74 & 128 & 184 \\\hline
DNN-LP-FBCOV &92.3 & 91.9 & 87.0 & 64 & 127 & 182  \\\hline 
DNN-QCP-FBCOV & {\bf 93.3} & {\bf 93.5} & {\bf 89.9} & {\bf 62} & {\bf 113} & {\bf 142} \\\hline

\multicolumn{7}{|c|}{Reference trackers} \\\hline
WSURF-1 & 86.6 & 82.7 & 80.8 & 87 & 223 & 228 \\\hline
KARMA &91.5 &89.4 &74.7 & {\bf 62} & 146 & 250 \\\hline 
Deep Formants & 91.7 & 92.3 & { 89.7} & 85 & 120 & { 143} \\\hline
\end{tabular}
}
\end{table*}
\subsection{DP-based formant tracking}

The FDRs (within $\tau_p=$~30\% and $\tau_a=$~300 Hz deviation) and FEEs for the different formant estimation methods are given in Table~\ref{tab:ftrack1}. 
In this table, the two metrics are computed 
by associating the three hypothesized formant tracks with the lowest three reference tracks.

The scores in the parentheses, however, denote the best scores which were obtained by identifying each formant as the spectral peak (among the detected five candidates) that was closest to the corresponding reference formant. The scores in the parentheses describe the performance of the underlying formant estimation method when used with an ideal formant tracker.
It can be seen from the results that the DP-based tracker gave scores inferior to the detection potential of the underlying spectral estimates.

It can be seen from Table~\ref{tab:ftrack1} that the QCP-based methods performed consistently better than all of their LP-based counterparts.
The covariance method performed better than the autocorrelation method for both LP and QCP. However, LP-FBCOV showed no improvement over LP-COV, and QCP-FBCOV seems to be inferior to QCP-COV, despite the detection potential being highest (the scores in parentheses) for QCP-FBCOV.
This behavior can be attributed to the inherent limitations of the DP-based tracker with a possibility of tracking a spurious candidate instead of the best candidate (which is otherwise not known without the ground truth).

\subsection{DNN-based formant tracking}

A comparison of the performance of the DNN-based formant tracker using LP-FBCOV and QCP-FBCOV for refinement is given in Table~\ref{tab:ftrack3} along with the performance of the DNN predictor. 

Three reference trackers (Wavesurfer (WSURF-1), KARMA, and Deep Formants) were chosen for comparison based on their performance shown in Table~\ref{tab:ftrack1}.    
It can be seen that the DNN-QCP-FBCOV tracker performed best, almost realizing the full potential of the QCP-FBCOV method (the scores in parentheses in Table~\ref{tab:ftrack1}). The improvement given by DNN-QCP-FBCOV compared particularly to the popular Wavesurfer tracker is large
showing a reduction of 29\%, 48\% and 35\% in the estimation error for the lowest three formants, respectively. These results demonstrate that the QCP-FBCOV method can be a good replacement for the popularly used LP-COV analysis in formant estimation and tracking tools and applications.

{
A detailed comparison in the formant tracking performance of KARMA, Deep Formants, DNN and DNN-QCP-FBCOV is given in Table~\ref{tab:ftrack1sound} for different phonetic categories (vowels, dipthongs and semovowels).
It can be seen that the proposed DNN-QCP-FBCOV method performed clearly better for all the phonetic categories in both FDR and FEE.  
Formant tracking performance of different methods analyzed separately for male and female speakers is given in Table~\ref{tab:gendercompare}. From the table it can be observed that the performance of Deep Formants is better for male speakers (except in $\delta F_1$, where the DNN-QCP-FBCOV method is better) but for female speakers the DNN-QCP-FBCOV method is better.

\begin{table*}[h]
\centering
\caption{\label{tab:ftrack1sound} The formant tracking performance of KARMA, Deep Formants, DNN and DNN-QCP-FBCOV in terms of FDR and FEE for different phonetic categories. 
The results are reported by averaging over of all the 192 utterances of the VTR test database.}
\resizebox{13cm}{8cm}{
\begin{tabular}{|c||c|c|c||c|c|c|}
\hline
 & \multicolumn{3}{c||}{FDR (\%)} & \multicolumn{3}{c|}{FEE (Hz)} \\\cline{2-7}
Phonetic category~ & ~$F_1$~ & ~$F_2$~ & ~$F_3$~ & ~$\delta F_1$~ & ~$\delta F_2$~ & ~$\delta F_3$\\\hline
\multicolumn{7}{c}{}\\\hline
\multicolumn{7}{|c|}{{\bf KARMA}} \\\cline{2-7} \hline 
Vowels (V)     & { 92.6} & 89.0 & 74.5 & {\bf 57.1} & 149.5 & 251.1  \\\hline
Diphthongs (D) & { 92.5} & 92.3 & 76.5 & {62.8} & 128.7 & 239.8  \\\hline
Semivowels (S) & { 86.9} & 86.9 & 73.6 & {\bf 76.1} & 154.8 & 258.3   \\\hline \hline
V+D+S      & { 91.5} & 89.4 & 74.7 & { 61.9} & 145.8 & 250.3  \\\hline
\multicolumn{7}{c}{}\\\hline

 \multicolumn{7}{|c|}{{\bf Deep Formants}} \\\cline{2-7} \hline
 Vowels (V)     & {92.7} & 93.7 & {\bf91.0} & 81.5 & 112.9 & {\bf135.4}  \\\hline
Diphthongs (D) & {93.2} & 93.8 & 90.6 & 84.8 & 112.2 & 132.9  \\\hline
Semivowels (S) & {87.0} & 86.1 & 84.4 & 96.1 & 148.4 & {\bf176.2}  \\\hline \hline
V+D+S      & {91.7} & 92.3 & 89.7 & 85.1 & 119.6 & 142.8  \\\hline
\multicolumn{7}{c}{}\\\hline
 \multicolumn{7}{|c|}{{\bf DNN}} \\\cline{2-7} \hline
Vowels (V)  &91.6 & 93.3 & 82.9 & 70.0 & 121.5 & 176.5  \\\hline
Diphthongs (D)  &92.9 & 92.6 & 84.6 & 71.5 & 120.9 & 175.2 \\\hline
Semivowels (S)  &84.5 & 85.2 & 76.6 & 89.2 & 155.9 & 214.5  \\\hline \hline
V+D+S      &90.5 & 91.6 & 82.1 & 73.9 & 127.9 & 183.6  \\\hline
\multicolumn{7}{c}{}\\\hline
 \multicolumn{7}{|c|}{{\bf DNN-QCP-FBCOV}} \\\cline{2-7} \hline
Vowels (V)  & {\bf94.2} & {\bf94.2} & 90.6 & {\bf57.1} & {\bf105.2} & 136.2   \\\hline
Diphthongs (D) & {\bf94.5} & {\bf95.2} & {\bf93.0} & {\bf61.9} & {\bf109.4} & {\bf119.5}  \\\hline
Semivowels (S) & {\bf88.8} & {\bf89.2} & {\bf84.6} & 76.6 & {\bf143.1} & 183.8  \\\hline \hline
V+D+S      & {\bf93.3} & {\bf93.5} & {\bf89.9} & {\bf61.7} & {\bf113.3} & {\bf141.8} \\\hline
\end{tabular}}
\end{table*}

\begin{table*}
\centering
\caption{\label{tab:gendercompare} Formant tracking performance of different methods for male and female speakers separately. The results are reported by averaging over all the utterances of the male and female speakers of the VTR test database.}
\vspace{-0.2cm}
\resizebox{13cm}{3.8cm}{
\begin{tabular}{|c||c|c|c||c|c|c|}
\hline
 & \multicolumn{3}{|c||}{FDR (\%)} & \multicolumn{3}{|c|}{FEE (Hz)} \\\cline{2-7}
Method & $F_1$ & $F_2$ & $F_3$ & $\delta F_1$ & $\delta F_2$ & $\delta F_3$\\\hline

\multicolumn{7}{c}{}\\\hline
\multicolumn{7}{|c|}{{\bf Male}} \\\hline
Deep Formants        &93.1 & 96.1 & 93.9 & 76 & 97 & 115   \\\hline

DNN    &89.4 & 91.8 & 83.3 & 75 & 126 & 177  \\\hline

DNN-QCP-FBCOV    & 92.6 & 94.3 & 90.5 & 60 & 109 & 137 \\\hline
\multicolumn{7}{|c|}{{\bf Female}} \\\hline
Deep Formants        &94.0 & 94.1 & 87.0 & 93 & 110 & 163  \\\hline
DNN     &92.7 & 91.0 & 79.6 & 72 & 133 & 196 \\\hline

DNN-QCP-FBCOV     & 94.5 & 91.9 & 88.8 & 65 & 123 & 151 \\\hline

\end{tabular}
}
\end{table*}

Formant tracking performance for different methods using speech degraded with white and babble noise at signal-to-noise ratio (SNR) levels of 10 dB and 5 dB are given in Table~\ref{tab:noisecompare}. From the table it can be observed that the proposed DNN-QCP-FBCOV method performed better in the case of speech degraded with white noise. In the case of speech degraded with babble noise, Deep Formants and DNN methods seems to perform better.


An illustration of formant tracking by KARMA and DNN-QCP-FBCOV for an utterance produced by a male speaker is shown in Fig.~\ref{fig:tvqcpvskarma}. It can be seen from the figure that the formants tracked by the DNN-QCP-FBCOV method match closely the ground truth in voiced segments. Furthermore, it can be clearly seen that DNN-QCP-FBCOV is better than KARMA in tracking all the formants. 

\begin{figure*}
\label{fig:tvqcpvskarma}
\includegraphics[width=17cm,height=12cm]{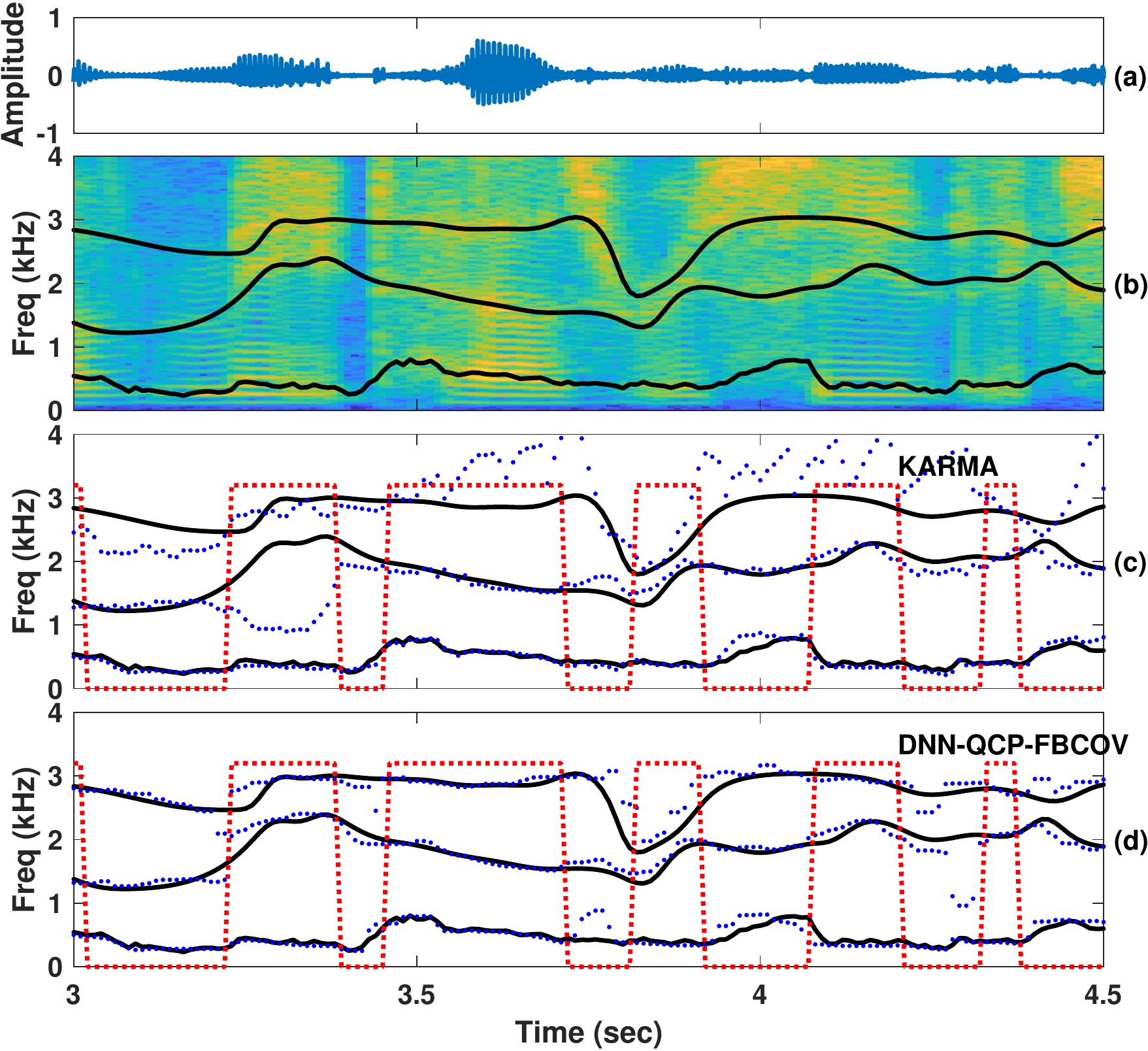} 
\vspace{-0.3cm}
\caption{Formant tracking by KARMA and DNN-QCP-FBCOV for an utterance produced by a male speaker: (a) the time-domain speech signal, (b) the narrowband spectrogram with reference ground truth formant contours, (c) the formant track estimates of KARMA along with the voiced-unvoiced regions shown by a dotted rectangular-wave plot, and (d) the formant track estimates of DNN-QCP-FBCOV.}
\end{figure*}

\begin{table*}
\centering
\caption{\label{tab:noisecompare} Formant tracking performance for different methods using speech degraded with white and babble noise at SNR levels of 10 dB and 5 dB. The results are reported by averaging over of all the 192 utterances of the VTR test database.}
\vspace{-0.2cm}
\resizebox{13cm}{7.5cm}{
\begin{tabular}{|c||c|c|c||c|c|c|}
\hline
 & \multicolumn{3}{|c||}{FDR (\%)} & \multicolumn{3}{|c|}{FEE (Hz)} \\\cline{2-7}
Method & $F_1$ & $F_2$ & $F_3$ & $\delta F_1$ & $\delta F_2$ & $\delta F_3$\\\hline




\multicolumn{7}{c}{}\\\hline

\multicolumn{7}{|c|}{{\bf White at 10 dB}} \\\hline
KARMA        &86.2 & 80.1 & 68.8 & 75.5 & 191.3 & 256.5 \\\hline
Deep Formants        &89.8 & 80.8 & {71.6} & 99.2 & 184.3 & {\bf238.7}   \\\hline
DNN &84.6 & 79.6 & 69.5 & 88.5 & 193.4 & 245.6    \\\hline

DNN-QCP-FBCOV     &{\bf91.1} & {\bf86.1} & {\bf71.8} & {\bf69.0} & {\bf162.2} & 251.8 \\\hline

\multicolumn{7}{|c|}{{\bf White at 5 dB}} \\\hline
KARMA        &80.1 & 72.5 & 64.0 & {91.6} & 232.5 & 279.2\\\hline
Deep Formants        &{\bf89.2} & 71.7 & {64.5} & 101.1 & 238.7 & {274.3}  \\\hline
DNN  &84.8 & 79.6 & {\bf69.1} & 88.1 & {\bf193.3} & {\bf247.4}   \\\hline

DNN-QCP-FBCOV     & 87.8 & {\bf80.6} & {65.3} & {\bf83.7} & {\bf196.9} & 282.8 \\\hline

\multicolumn{7}{c}{}\\\hline

\multicolumn{7}{|c|}{{\bf Babble at 10 dB}} \\\hline
KARMA        &90.3 & 83.8 & 71.8 & {\bf65.1} & 176.1 & 246.0  \\\hline
Deep Formants        &{\bf91.1} & 86.6 & {\bf81.7} & 88.4 & {145.9} & {\bf182.7}   \\\hline
DNN & 88.8 & {\bf89.1} & 78.8 & 77.7 & {\bf141.7} & 198.7   \\\hline
DNN-QCP-FBCOV    &90.7 & {87.1} & 81.1 & {66.0} & 153.9 & 203.8 \\\hline

\multicolumn{7}{|c|}{{\bf Babble at 5 dB}} \\\hline
KARMA        &88.2 & 78.9 & 68.7 & {\bf70.9} & 200.9 & 260.3   \\\hline
Deep Formants        &{\bf89.8} & 81.4 & {76.1} & 89.9 & {177.3} & {\bf209.1}   \\\hline
DNN  &87.5 & {\bf86.5} & {\bf76.2} & 80.5 & {\bf155.1} & 211.0   \\\hline

DNN-QCP-FBCOV     &87.7 & {81.7} & 74.9 & 72.4 & 187.9 & 239.7 \\\hline



\end{tabular}}
\end{table*}}

\section{Conclusions}
\label{sec:summary}
Formant tracking was studied in this paper based on the widely used two-stage approach consisting of the estimation stage and the tracking stage. In the former, six different all-pole modeling methods were first compared with a DP-based tracker. In addition, five known formant trackers were used as references. 
Two most potential all-pole modeling methods (LP-FBCOV and QCP-FBCOV) were then used with a modern DNN-based tracker by proposing a novel formant tracking technique which combines benefits of data-driven and model-driven approaches: the formants predicted with the data-driven DNN were refined using the frequencies of the peaks in the all-pole spectra computed by the model-driven LP-FBCOV and QCP-FBCOV methods. 


The DNN-based formant trackers using the LP-FBCOV and QCP-FBCOV refinement were further compared to two conventional formant trackers (Wavesurfer and KARMA) and to one recently published DNN-based tracker (Deep Formants). With the QCP-FBCOV refinement, the DNN-based tracker outperformed the conventional reference formant trackers in all metrics. Compared to Deep Formants, the proposed DNN-tracker gave better performance in all other metrics except for FDR and FEE in $F_3$ where Deep Formants was just slightly better. In addition to these encouraging objective results, it is worth emphasising that the proposed QCP-FBCOV refinement technique can be used in principle to improve the performance of any existing DNN-based formant tracker which has been trained to map a speech signal frame into formants, that is, there is no need to re-train the DNN-based tracker used. However, it should be noted that the performance of the proposed method might depend on the accuracy of the estimated glottal closure instants, which are needed to generate the QCP weighting function \cite{Manu2014}. { Therefore, the robustness of the proposed method in various noisy conditions needs to be studied further. {One way of improving the performance under degradations could be by training the DNN models for all noisy conditions of interest.}
Nevertheless, under clean conditions, the current study shows that the QCP-FBCOV method is a potential all-pole modeling technique to be used in formant tracking instead of the widely used conventional LP methods. }

{It is worth noting that the DNN-based formant tracking methods studied in this investigation (i.e. Deep Formants, DNN, DNN-LP-FBCOV, and DNN-QCP-FBCOV) are based on supervised learning and their computational complexity is relatively high compared to traditional model-based approaches. It is known that DNNs are resource hungry due to their need for training data and the architecture of the neural network adds more computational complexity when the trained network is used in formant tracking. }
{The LSTM-based Deep Formants architecture has approximately 4M parameters, while the FFNN DNN architecture we propose has around 0.3M parameters.
It is worth emphasizing, however, that the main contribution of this study, the QCP-FBCOV based refinement of formants, can be plugged into any existing pre-trained DNN-based tracker, which results only in a marginal increase in complexity.
Compared to the conventional autocorrelation based LP, which has a computational complexity of $O{(n^2)}$, our proposed QCP-FBCOV-based tracker has $O{(n^3)}$ complexity, where $n$ denotes the size of the covariance matrix (which is equal to the LP order, which was $p=12$ in the experiments of the current study).
However, the order of LP analysis being small, our proposed DNN-QCP-FBCOV-based formant tracking results only in a negligible increase in the overall computational complexity.
There is an added computation complexity due to the computation of the temporal weighting function ($w_n$ in Eqs.~\ref{eqn:wtfn1}~and~~\ref{eqn:wtfn2}), which calls for estimating glottal closure instants, which also requires LP inverse filtering and is proportional to $O{(n^3)}$ computations.}

\bibliographystyle{IEEEtran}
\bibliography{strings,refs}


\begin{biography}[{\includegraphics[width=1.1in,height=1.25in,clip,keepaspectratio]{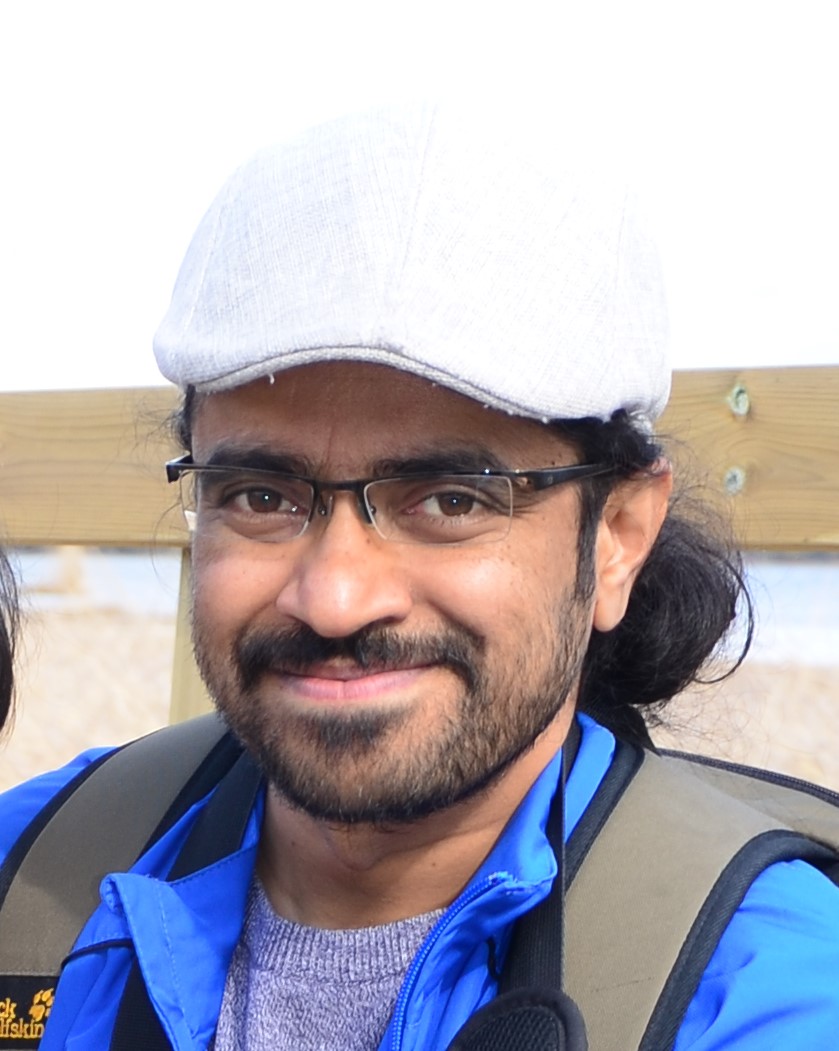}}]{Dhananjaya N. Gowda } is currently working as a Principal Engineer at Speech Processing Lab, AI Center, Samsung Research at Seoul R\&D Campus, South Korea. He worked as a postdoctoral researcher at Aalto University, Espoo, Finland from 2012 to 2017. He holds a Doctorate (2011) and a Master's degree (2004) in the area of speech signal processing, both from the Dept. of Computer Science and Engineering, Indian Institute of Technology (IIT) Madras, Chennai, India. His research interests include speech processing, signal processing, speech recognition, machine learning and spoken language understanding.
\end{biography}

\begin{biography}[{\includegraphics[width=1.1in,height=1.25in,clip,keepaspectratio]{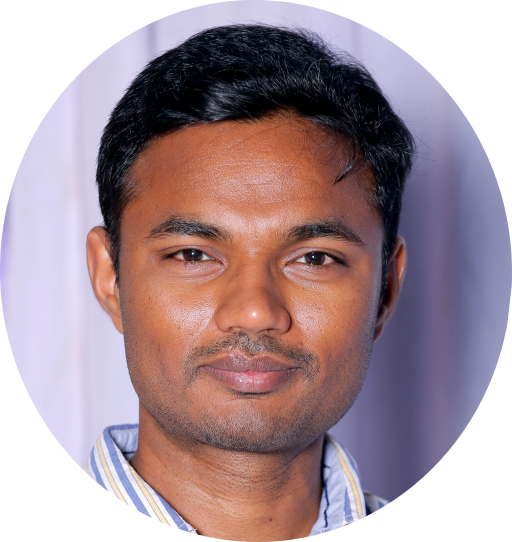}}]{Bajibabu Bollepalli} is currently working as an Applied Scientist at Amazon, Alexa AI, Cambridge UK. He worked as a Post-doctoral Fellow in Verisk Analytics, Munich, Germany before joining in Amazon. He received Ph.D degree from the Department of Signal Processing and Acoustics, Aalto University, Finland in 2020. He holds  Licentiate of Engineering (2017) from KTH Royal Institute of Technology, Stockholm, Sweden, and Master's (2012) and Bachelor's (2011) degrees both from IIIT-Hyderabad, India. He is a recipient of two best student paper awards one in ICASSP 2016 and another in Interspeech 2016. His research interests include speech synthesis, speech processing, natural language processing and machine learning.

\end{biography}

\begin{biography}[{\includegraphics[width=1.1in,height=1.3in,clip]{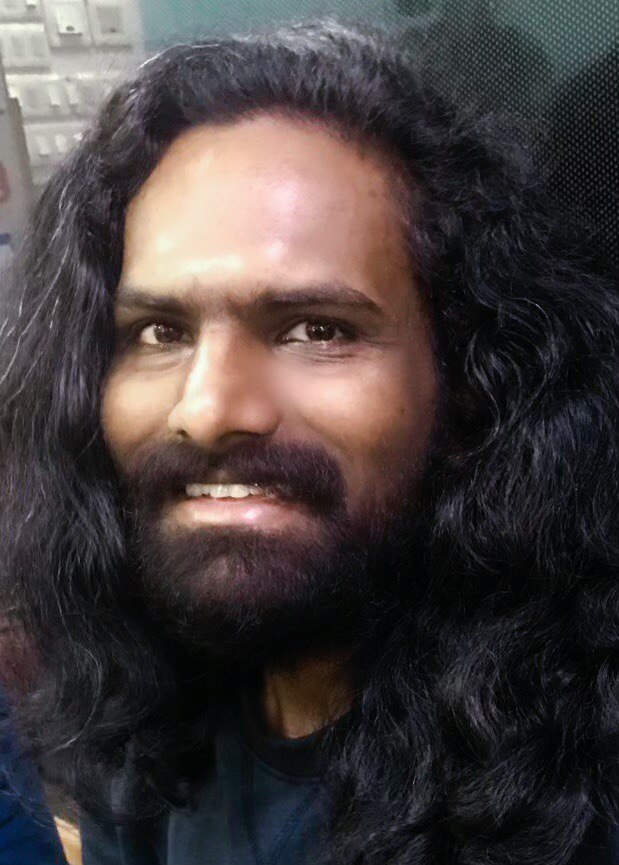}}]{Sudarsana Reddy Kadiri} received the B.Tech. degree from Jawaharlal Nehru Technological University (JNTU), Hyderabad, India, in 2011, with a specialization in electronics and communication engineering (ECE), the M.S. (Research) during 2011-2014, and later converted to Ph.D., and received Ph.D. degree from the Department of ECE, International Institute of Information Technology, Hyderabad (IIIT-H), India in 2018. He was awarded the Tata Consultancy Services (TCS) fellowship for his Ph.D. He is currently a Postdoctoral Researcher with the Department of Signal Processing and Acoustics, Aalto University, Espoo, Finland.  He was a Teaching Assistant for several courses at IIIT-H, from 2012 to 2018, and he has been involving in teaching and mentoring activities at Aalto University, since 2019. His research interests include signal processing, speech analysis, speech synthesis, paralinguistics, affective computing, voice pathologies, machine learning, and auditory neuroscience. He has published over 50 research papers in peer-reviewed journals and conferences in these areas. 
\end{biography}

\begin{biography}[{\includegraphics[height=1.25in,trim=0cm 0 0cm 0,clip]{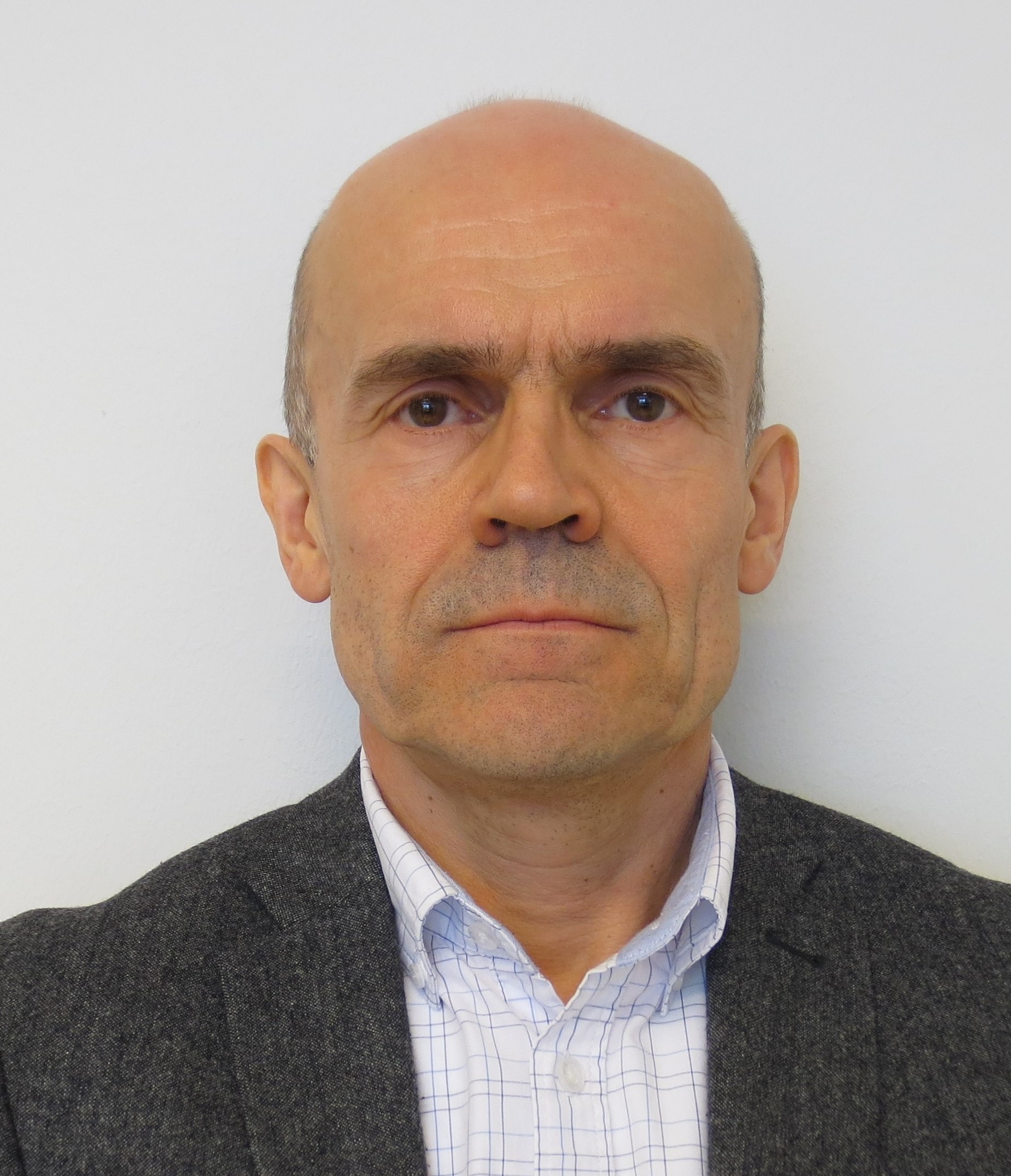}}]{Paavo Alku} received his M.Sc., Lic.Tech., and Dr.Sc.(Tech) degrees from Helsinki University of Technology, Espoo, Finland, in 1986, 1988, and 1992, respectively. He was an assistant professor at the Asian Institute of Technology, Bangkok, Thailand, in 1993, and an assistant professor and professor at the University of Turku, Finland, from 1994 to 1999. He is currently a professor of speech communication technology at Aalto University, Espoo, Finland. His research interests include analysis and parameterization of speech production, statistical parametric speech synthesis, spectral modelling of speech, speech-based biomarking of human health, and cerebral processing of speech. He has published around 220 peer-reviewed journal articles and around 210 peer-reviewed conference papers. He is an Associate Editor of J. Acoust. Soc. Am. He served as an Academy Professor assigned by the Academy of Finland in 2015-2019. He is a Fellow of the IEEE and a Fellow of ISCA. 
\end{biography}
\EOD
\end{document}